\documentclass{jpsj2}

\usepackage[dvips]{color} 

\newcommand{\infinity}{\infty}

\title{ Modeling of the Peeling Process of Pressure-sensitive Adhesive Tapes with the
Combination of Maxwell Elements }

\author{Katsuhiko \textsc{Sato}$^{1}$\thanks{ E-mail address: sato@complex.c.u-tokyo.ac.jp
} and Akihiko \textsc{Toda}$^{2}$\thanks{E-mail address: atoda@hiroshima-u.ac.jp} }

\inst{$^{1}$Department of Pure and Applied Sciences, University of Tokyo, Komaba,
     Meguro-ku, Tokyo 153-8902, Japan \\ $^{2}$ Faculty of Integrated Arts and Sciences,
     Hiroshima University, Higashi-Hiroshima 739-8521, Japan }

\abst{A simple model for the peeling process of pressure-sensitive adhesive tape is
presented.  The model consists of linear springs and dashpots and can be solved
analytically. Based on the modeling, the curved profile of the peeling tape is
spontaneously determined in terms of viscoelastic properties of adhesives.  Using this
model, two experimental results are discussed: critical peel rates in the peel force and
the peel rate dependence of the detachment process of adhesive from the substrate.  }

\kword{adhesive, Maxwell element, critical rates}

\begin{document}
\maketitle

\section{Introduction}\label{sec:intro}

The behaviors of the peeling process of pressure-sensitive adhesive tape are very complex
and interesting: (i) The peel force curve against the peel rate sometimes has two or three
local maximums (Fig.~\ref{fig:01})~\cite{Handbook1989,Kaelble1969}. This fact implies that
we will observe self-excited oscillation of the peel force, if the adhesive tape is peeled
at a rate in the negative slope in Fig.~\ref{fig:01}~\cite{Handbook1989,Yamazaki2002}. (ii)
We observe fingerlike instabilities at the peel line, which is the interface between the
adhesive and the rigid substrate (Fig.~\ref{fig:02}), and the front of failure waves along
the peel line~\cite{Ashby1976}.  (iii) In some adhesives, we observe caves inside the
adhesive~\cite{Urahama1989}.  The appearance of the caves indicates the flow of the
adhesive; the adhesives are polymer melts with viscoelasticity.

In this paper, in order to describe the peeling system with viscoelastic adhesives, we
present a simple model made of linear strings and dashpots and solve it analytically.
A few attempts have already been made to incorporate the viscoelastic
properties~\cite{Aubrey1984,Mizumachi1985}.  The advantage of our modeling is on the
spontaneous determination of the curved profile of the peeling tape with the
viscoelastic adhesives.

Taking the advantage, we shall discuss the followings: (i) "critical peel rate" at which
the peel force curve takes a local maximum and (ii) the positive dependence of the
peel force on the peel rate, which is not trivial. Actually, it is not
straightforward to derive positive dependence, and we approach this difficulty by
considering more microscopic picture at the interface between the adhesive and the
substrate in $\S$~\ref{sec:positive}.

\section{Model} \label{sec:model}

\subsection{Frame of the model}

\begin{figure}[tb]
\begin{center} 
\includegraphics[width=12cm]{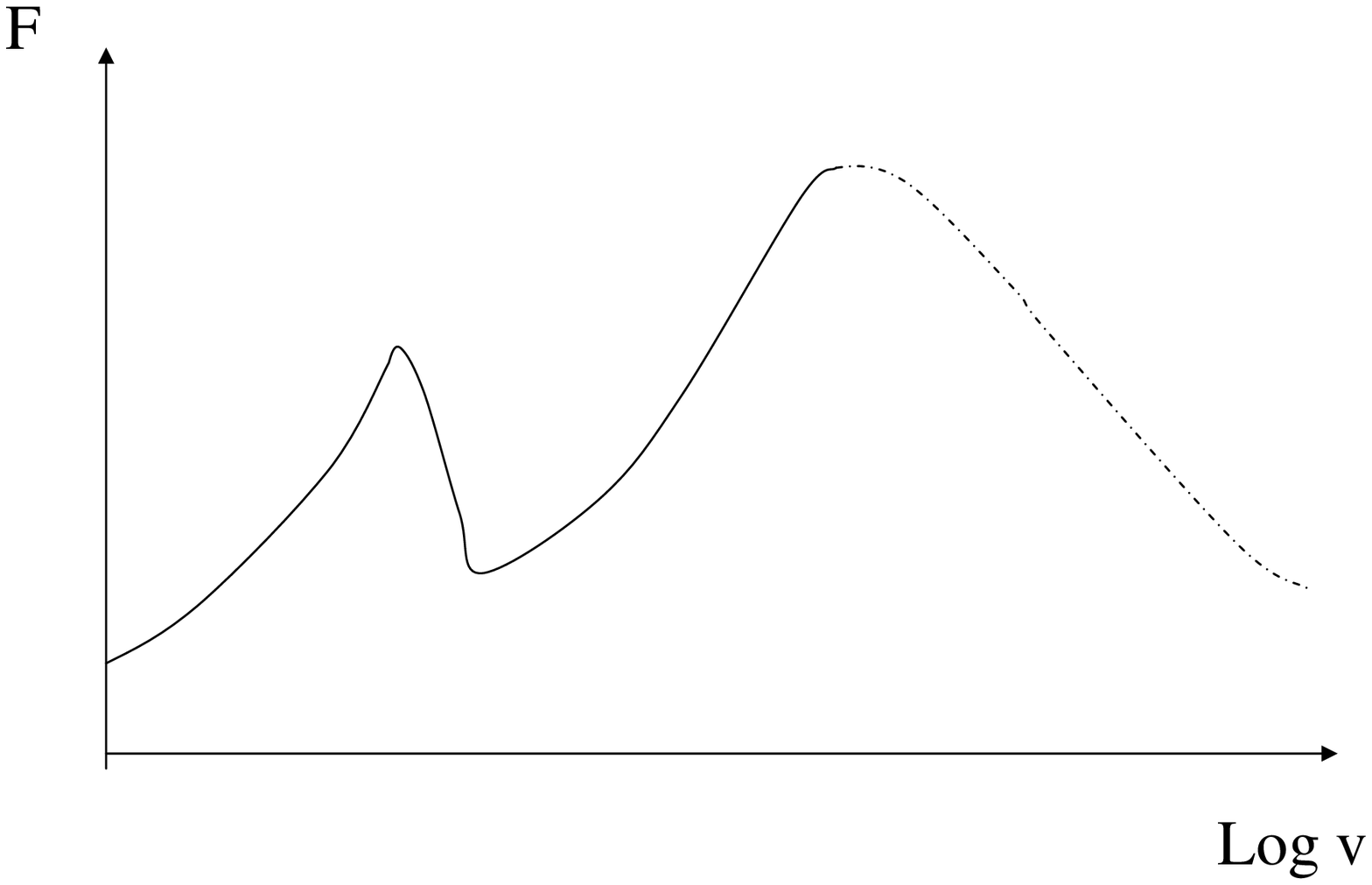}
\end{center} 
\caption{Schematic picture of the peel force against peel rate observed experimentally. 
The solid line indicates the peel force when the system is in the steady state, while the
dashed line indicates the average value of the peel force; oscillation never stops in that
rate region.  } \label{fig:01}
\begin{center} 
\includegraphics[width=12cm]{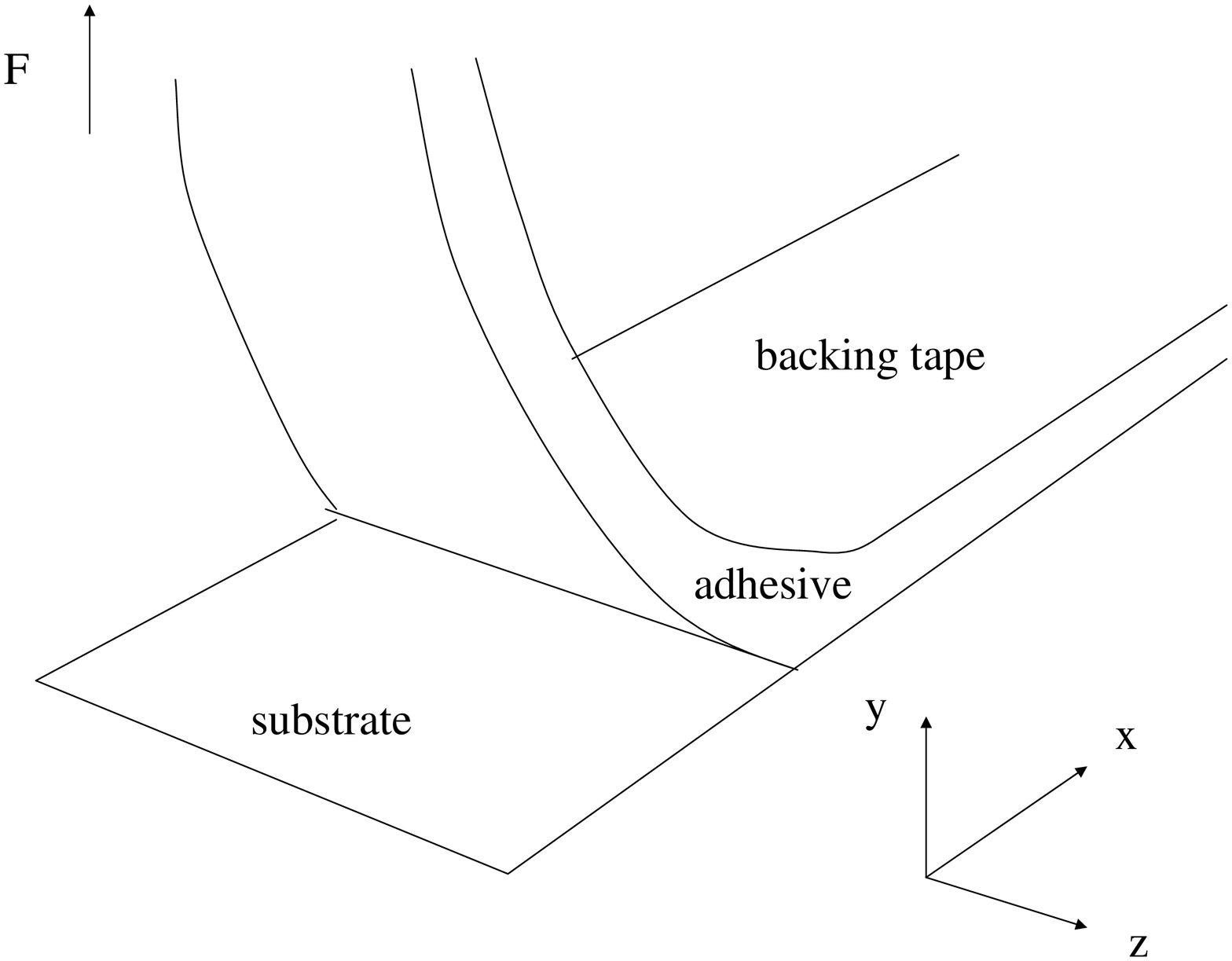}
\end{center} 
\caption{Schematic illustration of the experimental system. 
} \label{fig:02}
\end{figure}

We shall consider the system in two dimensions of the x-y plane in Fig.~\ref{fig:02}. In
practice, there is z-direction dependence of the system with fingerlike instabilities
along the peel line.  However, with small amplitude of the fingering, this assumption will
be justified.

\begin{figure}[hbtp]\begin{minipage}[t]{14cm} \begin{center}
\includegraphics[width=12cm]{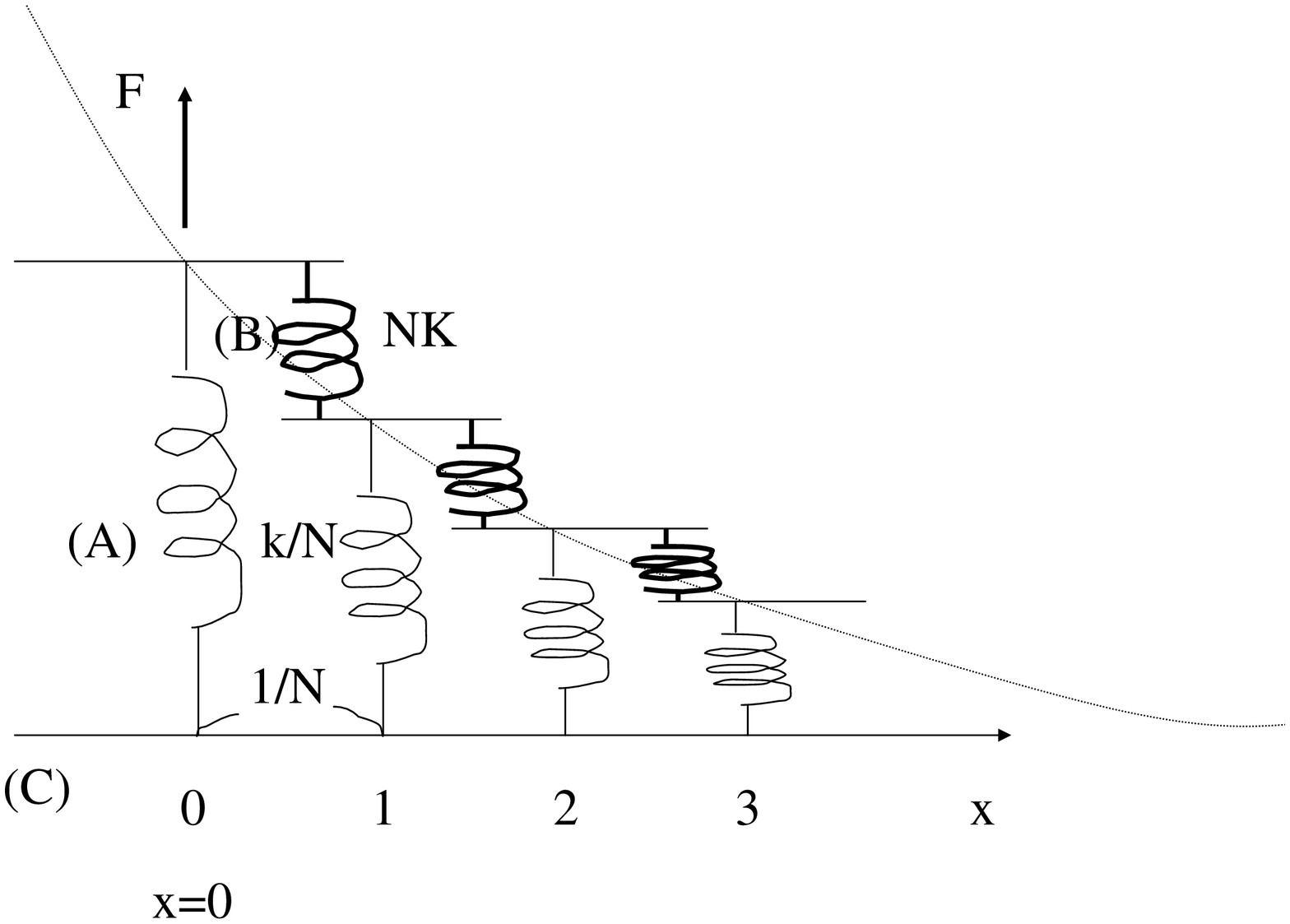} \caption{Schematic picuture of the
 prototype of the model with spring elements. }
\label{fig:03} \end{center} \end{minipage}

\begin{minipage}[t]{14cm} \begin{center} 
\includegraphics[width=12cm]{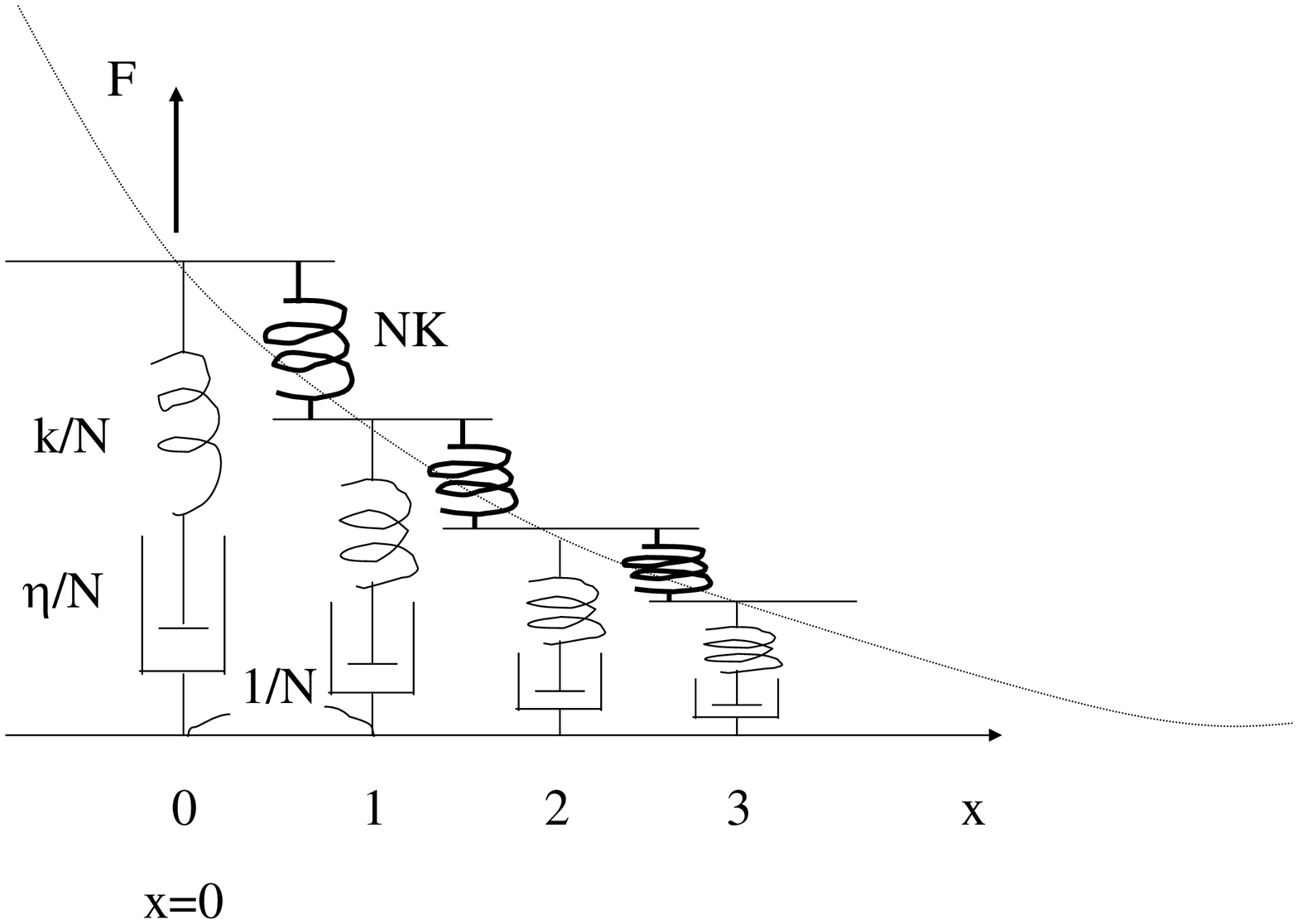}
\caption{Schematic picuture of the model in the case of Maxwell element.  } 
\label{fig:04} \end{center}
\end{minipage}\end{figure}

The adhesive is represented by a set of linear springs (A) with spring constant $k/N$
where $N$ is the number of the springs per unit length along the $x$ axis. These springs
are numbered along the $x$ axis as indicated in Fig.~\ref{fig:03}; $k$ corresponds to the
modulus of elasticity of the adhesive $E_a$ and will be represented as $k= E_a \omega /
y_0$ for the adhesive with the width $\omega$ and initial thickness $y_0$. These springs
can be elongated only along the y axis. We shall call this part of the model just
"element"; the element will be replaced by another element, such as a series connection
of linear spring and dashpot (the so-called Maxwell element) or various
linear combinations of linear springs and dashpots.

Secondly, corresponding to the backing tape shown in Fig.~\ref{fig:02}, we consider
flexible joints which connect adjoining springs (A) and are represented by linear springs
(B) with spring constant $N K$, as shown in Fig.~\ref{fig:03}.  This modeling assumes
that the backing tape is a thin elastic object without bending; $K$ corresponds to the
modulus of elasticity of the backing tape $E_b$ and will be represented as $K= E_b \omega
\delta$ for the backing tape with the width $\omega$ and thickness $\delta$.  We admit
that it is more realistic to describe the backing tape as a beam with flexural rigidity~\cite{Bikerman1957}. However, if we adopt the bending of the beam, the equation
system becomes much more complex, as shown in Appendix\ref{appen:A}, and the results are
not essentially different from the results with the flexible joint of the linear springs
(B).  Therefore, we adopt this simpler model of the linear springs (B) as the flexible
joint in the main part of the following discussion.  In terms of the rigid substrate, it
is represented by a horizontal line (C) (the x axis) as in Fig.~\ref{fig:02}. It is
assumed that all elements and springs (B) are massless.

We impose on the elements a condition that the bottom of each element is detached from the
line (C) when the force applied to the element becomes some critical value
$\sigma_c/N$. Imposing this condition on the element corresponds to the situation where
the adhesive is detached from the substrate when the normal stress of the adhesive at the
interface reaches the critical value $\sigma_c/w$. Until $\S$~\ref{sec:positive}, we
assume that the quantity $\sigma_c$ is constant; the possibility of the peel rate
dependence of $\sigma_c$ is discussed in $\S$~\ref{sec:positive}. As a natural boundary
condition, we assume there is no force at $x=\infinity$.

Under these assumptions, let us now consider the following situation: the linear springs
(A) and (B) are settled in the range [$-\infinity$,$\infinity$] and there is no force
acting on the springs in the initial stage. At $t=-\infinity$, we begin to lift up
quasi-statically one spring (B) at the left end at $x=-\infinity$ ; since we move it
quasi-statically, the force balance holds for every element at every moment. We suppose
that the force applied to the element at $x=0$ reaches the critical value $\sigma_c/N$ at
$t=0$ (we call such an element the ``edge'' of the elements hereafter because it is the
boundary between the elements having left the substrate and the elements sticking to the
substrate).

Under such situation, let us now try to obtain the shape of the curve drawn by the top
ends of the elements (we call this curve "flexible joint's curve" for brevity).  First, we
consider the force balance equation at the $i$-th element ($i > 0$):
\begin{equation} -\frac{k}{N} f_i - N K (f_i-f_{i+1}) + N K (f_{i-1}-f_{i}) =0,
\label{eq:01} \end{equation} 
where $f_i$ is the y coordinate of the top end of the
$i$-th element. From the conditions at $i=0$ and $i=\infinity$, we obtain the
following relations: 
\begin{eqnarray} \frac{k}{N} f_0 &=& \frac{\sigma_c}{N}
\label{eq:02} \\  f_{\infinity} &=& 0 .  \label{eq:03} \end{eqnarray}
Taking the limit $N \rightarrow \infinity$ in these relations and rewriting the coordinate
$f_i$ by $f(x)$ through the relation $x=\frac{i}{N}$, we obtain the equation for the
flexible joint's curve and its boundary conditions from (\ref{eq:01})-(\ref{eq:03}):
{
\begin{eqnarray} 
K f''(x) &=& k f(x) \label{eq:04} \\ Kf''(0) &=& \sigma_c \label{eq:05} \\ f(\infinity)
 &=& 0 \label{eq:06}
\end{eqnarray} 
where the prime over the function $f$ means its derivative with respect to $x$.  } Solving
eq. (\ref{eq:04}) under the boundary conditions (\ref{eq:05}) and (\ref{eq:06}), we
have 
\begin{equation}
f(x)=\frac{\alpha^2 \sigma_c}{K} e^{- x / \alpha} \label{eq:07}, \end{equation} 
where $\alpha=\sqrt{\frac{K}{k}}$; $\alpha$ characterizes the range where the elements are
stretched (we call this length the "stretching length" for brevity).

From the force balance relation at the 0th element, the peel force $F$ applied to the
spring (B) at $x=0$ is represented by the flexible joint's curve $f$, as follows,
\begin{equation}
F = \frac{k}{N} f_0 + N K (f_0-f_1) .
\end{equation} 
Then, after taking the limit $N \rightarrow \infinity$, we have 
\begin{equation} F = -K f'(0)  \label{eq:16} \end{equation} 
which implies for this model that
\begin{equation} F = \alpha \sigma_c .  \label{eq:162} \end{equation} 

It is noted that the flexible joint's curve obtained in our modeling with more complex
elements comprised of linear combination of springs and dashpots is also of the
exponential form as
\begin{eqnarray} 
f(x)=f(0)\, e^{- x/\beta} , \label{eq:980}
\end{eqnarray}
where $f(0)$ and $\beta$ are some positive constants; the proof is given in Appendix\ref{appen:B}. 
From the boundary condition (\ref{eq:05}) which is satisfied with more complex elements, 
$f(0)$ is turned out to be $f(0)= \beta^2 \sigma_c /K$. 
Then, with the general relationship (\ref{eq:16}) which is also satisfied with more 
complex elements, the peel force always has the form of 
\begin{eqnarray} 
F= \beta \sigma_c . \label{eq:979}
\end{eqnarray}
This relation means that the peel force is given by the multiplication of the critical
stress $\sigma_c$ and the stretching length $\beta$.  In the present modeling, this
stretching length $\beta$ is spontaneously determined in terms of the elastic moduluses of
the adhesive, $k$, and of the flexible joint, $K$.  Therefore, when we keep both the
elastic modulus of the flexible joint $K$ and the critical stress $\sigma_c$ constant, the
peel force monotonically decreases as the adhesive gets stiffer with larger $k$.  This
relation becomes important in understanding the behavior with viscoelastic elements
showing $v$ dependence of $k$, as discussed in the following.  

\subsection{Case of Maxwell elements}

{As mentioned in $\S$~\ref{sec:intro}, since the adhesive is a polymer melt with
visco-elasticity, it is more plausible to describe the adhesive in terms of Maxwell
element of serial connection of spring and dashpot with spring constant, $k$, and
viscosity, $\eta$.  For the serial connection of spring and dashpot, the force acting on
the element is given as $ k y_1 = \eta \dot{y_2} , $ where $y_1$ and $y_2$ represent the
displacements of the spring and dashpot, and the dot over $y_2$ means its time derivative;
the total displacement $y$ is given as $ y = y_1 + y_2 .  $ The general solution of this
system gives the applied force as $ k \int _{-\infty} ^{t} e^{-\frac{t-t'}{\tau}}
\dot{y}(t') dt' $ , where $\tau=\eta/k$ is the relaxation time of the Maxwell element.

Suppose that the spiring (B) at the left end is lifted up with constant speed $v$ at
$t=-\infinity$ and the system is in the steady state, then the $y$ coordinate of the
$i$-th element sticking to the substrate will be expressed as $f(i/N - vt)$.  If the edge
of the elements reaches $x=0$ at $t=0$, we get the following equation of $f(x)$ for the
limit of $N \rightarrow \infinity$ from the force balance at the $i$-th element ($i > 0$):
\begin{eqnarray} K f''(x) \!\!\!&=&\!\!\!  k \int_{-\infinity}^{0}
e^{t'/\tau} \dot{f}(x-v t') dt', \label{eq:08} \end{eqnarray} where $x$ corresponds to
$i/N$. It is useful to note that the visco-elastic properties of the element appears only
in the right hand side of the above equation. The boundary conditions are the same as
eqs.~(\ref{eq:05}) and (\ref{eq:06}).  }

Suppose that the solution of this equation has an exponential form of eq.~(\ref{eq:980})
and substitute it into eq. (\ref{eq:08}), then we get an algebraic equation for
$\beta$:
\begin{equation} \beta^{2} - \frac{\alpha^2}{v \tau} \beta - \alpha^{2}=0 . 
\end{equation} 
where the condition $\beta > 0$ is set by the boundary condition
(\ref{eq:06}), and hence
\begin{equation} \beta = \frac{\alpha^2}{2 v \tau}+ \frac{1}{2}
\sqrt{(\frac{\alpha^2}{v \tau})^2+ 4\alpha^{2}}.  \label{eq:10} \end{equation} Using the
boundary condition (\ref{eq:05}), we get $f(x)= (\beta^2 \sigma_c / K) e^{- x / \beta}$
and we obtain the peel force represented as eq.~(\ref{eq:979}) from the relationship
(\ref{eq:16}).  We have plotted in Fig.~\ref{fig:12} the peel force $F$ against the peel
rate $v$ for $\tau=1$, $k=1$, $K=1$, and $\sigma_c=1$.

For the case of Maxwell element with viscoelasticity, the exponential deformation brings
an additional feature that the whole elements behave in a similar way; {\it i.e.} all
elements behave as a viscous liquid at low peel rate and as an elastic solid at high rate,
as shown in the following expression of the applied force to each element,
\begin{equation}
k \int_{-\infinity}^{0} e^{t'/\tau} \dot{f}(x-v t') dt' =
\frac{v\tau\alpha}{1+v\tau\alpha} k f(x) = \frac{1}{1+v\tau\alpha} \eta \dot{f}(x)
\end{equation}
In this sense, this system of peeling adhesive tape is essentially different from the
system of a fracture of adhesive studied by de Gennes~\cite{deGennes1996}, where the
behavior of an viscoelastic adhesive depends on the position in the system.  As the
consequence of this behavior at the limiting of $v\rightarrow 0$, the stretching length
$\beta$ expressed by eq.~(\ref{eq:10}) and consequently the peel force $F$ diverges at $v=
0$, as shown in Fig.~\ref{fig:12}.  The divergence of the peel force at $v=0$ is obviously
unrealistic.  In the next section, we introduce more realistic model with a combination of
springs and dashpots to describe the adhesive of polymer melt with slight cross-links.

{ We finally comment on the stability of this system.  In the present analysis, the
expression of $f(x - vt)$ assumes the instantaneous adjustment to the stationary state and
hence assumes the stability.  From the stability analysis of an analogous
situation~\cite{sato1999}, we can put one comment to justify the assumption.  If the whole
system is hard enough to transfer the change in applied load to the elements in a shorter
time interval in comparison with the relaxation time of the elements, the faster rate of
detachment of the elements immediately brings smaller applied force on the elements under
the condition of constant speed of lifting so that the detachment rate becomes slower, and
vice versa.  That is, for the hard system, the peeling edge tends to move with constant
speed equal to the lifting velocity.  }

\section{Explanation of Two Critical Peel Rates in Terms of the Model} \label{sec:two}

{ As mentioned above, the adhesive is a polymer melt with small portions of
cross-linking.  The visco-elastic properties of the response to the deformation can be
represented by two characteristic times; they are the times characterizing the transitions
among the glass-like response for fast deformations, the rubber-like responses with
entanglements behaving as temporal cross-links and with true cross-linking for
intermediate and slow rates of deformation, respectively, which are denoted by} { $\tau_g$
and $\tau_r$ ($\tau_g < \tau_r$), } respectively~\cite{Viscoelastic1970}.

\begin{figure}[hbtp]\begin{minipage}[t]{14cm} \begin{center} 
\includegraphics[width=12cm]{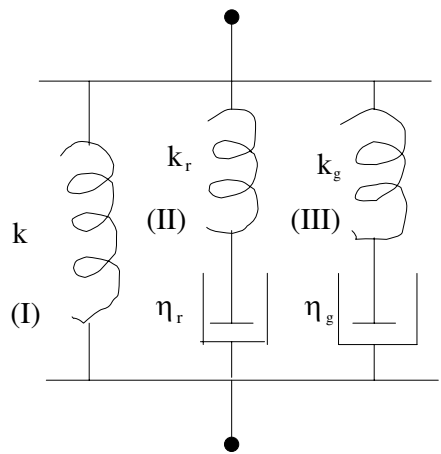}
\caption{ Schematic picuture of the element representing adhesive. The element has two
relaxation times and the elasticity due to cross-links within the adhesive.  }
\label{fig:05} \end{center}
\end{minipage}\end{figure}

Thus we should construct the element of the model so as to have two relaxation times, as
shown in Fig.~\ref{fig:05}, which consists of one linear spring with spring constant $k$
and two different Maxwell elements which are specified by the elastic constants, $k_r$ and
$k_g$, and viscous constants, $\eta_r$ and $\eta_g$, respectively; these quantities are
related to the relaxation times $\tau_r$ and $\tau_g$ as: $\tau_r=\eta_r/k_r$ and
$\tau_g=\eta_g/k_g$. We call each part of the element "spring (I)", "Maxwell element (II)"
and "Maxell element (III)," respectively.  Note that spring (I) represents the effect of
the cross-linked polymer within the adhesive.

The equation for $f(x)$ is 
\begin{eqnarray} f''(x) &=& \alpha^{-2} f(x) + {\alpha_r}^{-2}
\int_{-\infinity}^{0} e^{t'/\tau_r} \dot{f}(x-v t') dt' + {\alpha_g}^{-2}
\int_{-\infinity}^{0} e^{t'/\tau_r} \dot{f}(x-v t') dt' 
\end{eqnarray} 
where $\alpha=\sqrt{K/k}$, $\alpha_r=\sqrt{K/k_r}$, and $\alpha_g=\sqrt{K/k_g}$. { The boundary
conditions are the same as eqs.~(\ref{eq:05}) and (\ref{eq:06}). }

\begin{figure}[hbtp]\begin{minipage}[t]{14cm} \begin{center} 
\includegraphics[width=12cm]{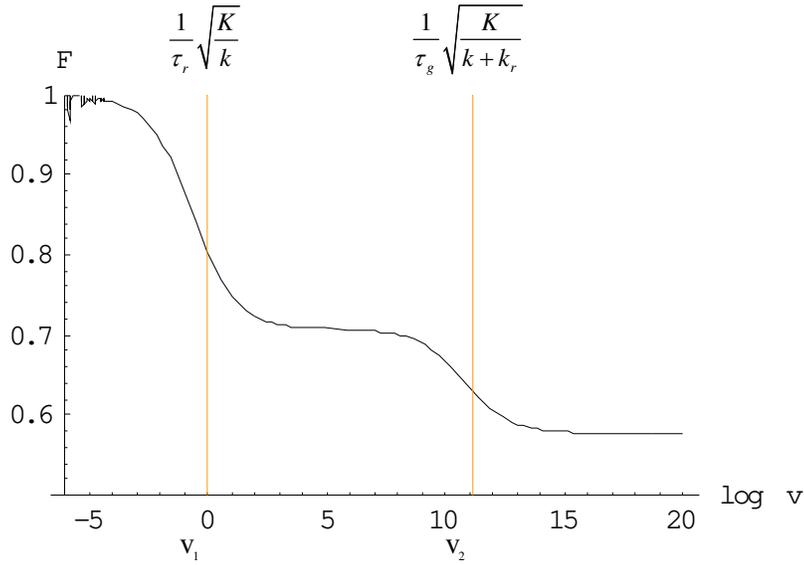}
\caption{ The plot of peel force against the logarithm of the peel rate calculated from
eqs.~(\ref{eq:979}) and (\ref{eq:998}).  Parameters are $\tau_g=0.00001$ , $\tau_r=1$ ,
$K=1$ , $k=1$ , $k_r=1$, $\sigma_c=1$, and $k_g=1$.  The gray line indicates $v_1$ and
$v_2$ evaluated from eqs.~(\ref{eq:11}) and (\ref{eq:12}). }
\label{fig:07} \end{center} \end{minipage}\end{figure}

The peel force is given as eq.~(\ref{eq:979}).  The stretching length $\beta$ is the
positive root of the following algebraic equation 
\begin{equation} 
\beta^{-2}= \alpha^{-2} +\alpha_r^{-2} v \tau_r /(\beta+v \tau_r) +\alpha_g^{-2} v \tau_g /(\beta+v \tau_g). 
\label{eq:998}
\end{equation}
Plotting the peel force $F$ against the logarithm of the peel rate $v$, as shown in
Fig.~\ref{fig:07}, we can find two critical peel rates labelled by $v_1$ and $v_2$.  These
two peel rates $v_1$ and $v_2$ corresponds to the rates at which the elastic modulus of
the element gradually changes from $k$ to $k+k_r$, and from $k+k_r$ to $k+k_r+k_g$,
respectively, with increasing peel rate.

At a very low rate, Maxwell elements (II) and (III) do not contribute to the elastic
modulus of the element because the elements have already been relaxed. Thus the
elastic modulus of the element turns out to be $k$ only, which is contributed by the
spring (I). From the result around (\ref{eq:07}), the ``stretching length'' in that case is
given by $\sqrt{K/k}$. Then, the time required for elongating each element until its
detachment from the substrate is estimated to be $(1/v) \sqrt{K/k}$, where $v$ is the
peel rate (we call the time "stretching time").  As the peel rate increases, the
stretching time $(1/v) \sqrt{K/k}$ gets closer to the relaxation time of Maxwell
element (II), $\tau_r$, and hence the elastic modulus of the element gradually
changes from $k$ to $k+k_r$.  Thus the peel rate $v_1$ is estimated as,
\begin{equation} v_1=\frac{1}{\tau_r} \sqrt{\frac{K}{k}} \label{eq:11} \end{equation}
By the same logic, we can also estimate the peel rate, $v_2$ as, 
\begin{equation}
v_2=\frac{1}{\tau_g} \sqrt{\frac{K}{k+k_r}} , \label{eq:12} \end{equation} 
In Fig.~\ref{fig:07}; we can actually see that the evaluated $v_1$ and $v_2$ correspond to
the boundaries of the ranges.

As shown in Fig.~\ref{fig:07}, in our model with constant critical stress for the
detachment $\sigma_c$, the peel force curve is a decreasing function of the peel rate
and does not reproduce the experimental results of the positive slope.  If $\sigma_c$
is an increasing function of the peel rate, the peel force curve will then recover
the positive dependence on peel rate with small regions having negative slopes around
the two critical rates $v_1$ and $v_2$.  Then, the peel force curve will become very
similar to the experimental results schematically shown in Fig.~\ref{fig:01}.

Of course, we cannot readily say the reason for the dependence of the critical normal
stress $\sigma_c$ on the peel rate (we are giving one possibility of the peel rate
dependence of the critical stress in the next section), but it is probably obvious
that the appearance of the two critical rates in the experiment is due to the changes
in elastic modulus of the adhesive.  Actually, by a lot of investigators, this
relationship between the change in elastic modulus of the adhesive and the critical
peel rates has been postulated~\cite{Handbook1989}. We have succeeded in showing the
relation between the critical peel rates and the relaxation times of the adhesive in
terms of the quantities characterizing the viscoelastic properties of the adhesives.

\section{On the Peel Rate Dependence of the Critical Normal Stress
$\sigma_c$}\label{sec:positive}

As discussed in the previous section, if we calculate the peel force in our model with
constant critical stress $\sigma_c$ for detachment, we inevitably observe negative
slope of the peel force, which is completely different from the experimental results
even in the qualitative sense.  Thus we must discuss the reason for the discrepancy
in the slope in more detail.

\begin{figure}[hbtp]\begin{minipage}[t]{14cm} \begin{center} 
\includegraphics[width=12cm]{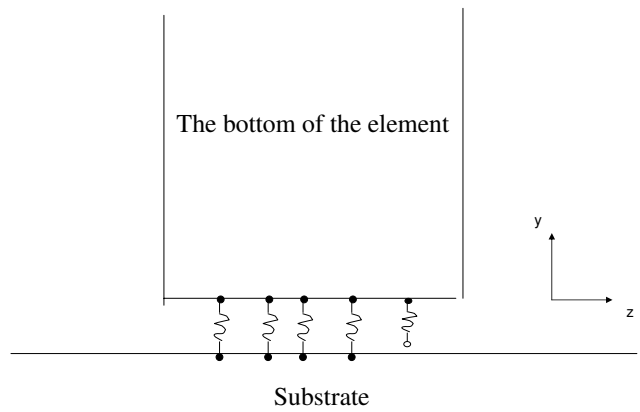}
\caption{ Enlarged schematic picture of the bottom of the element. We shall consider the
bottom of the element as an object consisting of small springs which are supposed to be
detached and attached to the substrate stochastically.  } \label{fig:08} \end{center}
\end{minipage}\end{figure}

The element of our model is a reduced one representing some part of the adhesive with
finite size, so that the element can be considered to consist of more microscopic
parts, such as polymer chains. So we can consider the bottom ends of the elements as
the assembly of more microscopic parts as illustrated in Fig.~\ref{fig:08} (we
call them ``small springs'').

Since these small springs are microscopic objects such as polymer chains, it is reasonable
to assume that their processes of detachment and attachment are stochastic; each
small spring is stochastically detached and attached to the substrate with some
probabilities which generally depend on the force applied to the small spring.

We expect that the small springs are detached from the substrate more often as the applied
force gets larger.  We may also expect that the attachment process of each small spring
depends on the applied force.  But the dependence will be weaker, because the detached
small springs will stay close to the substrate and will attach to the substrate almost
equally for finite applied force.  Thus we shall write down the time evolution equation
for the amount $A$ of small springs being attached to the substrate (which is a normalized
quantity) as: 
\begin{eqnarray} \frac{dA(t)}{dt} &=& - \frac{A(t)}{\tau_{-}(
\sigma(t),A(t))}+ \frac{1-A(t)}{\tau_+}, \label{eq:13} \\ &\equiv& G(\sigma(t),A(t)) \\
\tau_-(\sigma,A) &=& \tau_0 \exp [\frac{U- \ell \sigma/A}{k_{B} T}] \\ \tau_+ &=& \tau_0
\end{eqnarray} 
where $\sigma$ is the tensile force acting on the whole element; $\sigma/A$
is the force acting on each small spring, $\ell$ is a constant having the dimensions of
length, which corresponds to the width of the potential hole representing the attachment
of small springs to the substrate. $\tau_0$ is a constant with the dimension of time, $U$
is a quantity representing the depth of the potential minimum, $k_{B}$ is the Boltzmann
constant, and $T$ is the temperature of the system. When the quantity $A$ vanishes the
element leaves from the substrate.

{ We first examine the stationary case, i.e., where $\sigma$ is independent of time.
Solving the equation $G(\sigma,A)=0$ for $A$ and remembering that the sign of $\partial
G(\sigma,A)/ \partial A $ determines the linear stability of the solution, we readily find
that the eqution system has its reasonable solution only when $\sigma$ is smaller than
some critical value $\sigma_s$ determined by the two equations $G(\sigma_s,A)=0$ and
$\partial G(\sigma_s,A)/ \partial A = 0$; namely, $\sigma_s$ is given by the following
equation, 
\begin{equation} \sigma_s + \ln(\frac{\sigma_s}{k_B T}) = U - k_B T.
\end{equation} 
If $\sigma$ is smaller than $\sigma_s$, $A$ finally reaches a finite value
in the steady state and the element is kept sticking to the substrate. If $\sigma$ is
larger than $\sigma_s$, on the other hand, $A$ finally vanishes and the element is
detached from the substrate.  }

\begin{figure}[hbtp]\begin{minipage}[t]{14cm} \begin{center} 
\includegraphics[width=12cm]{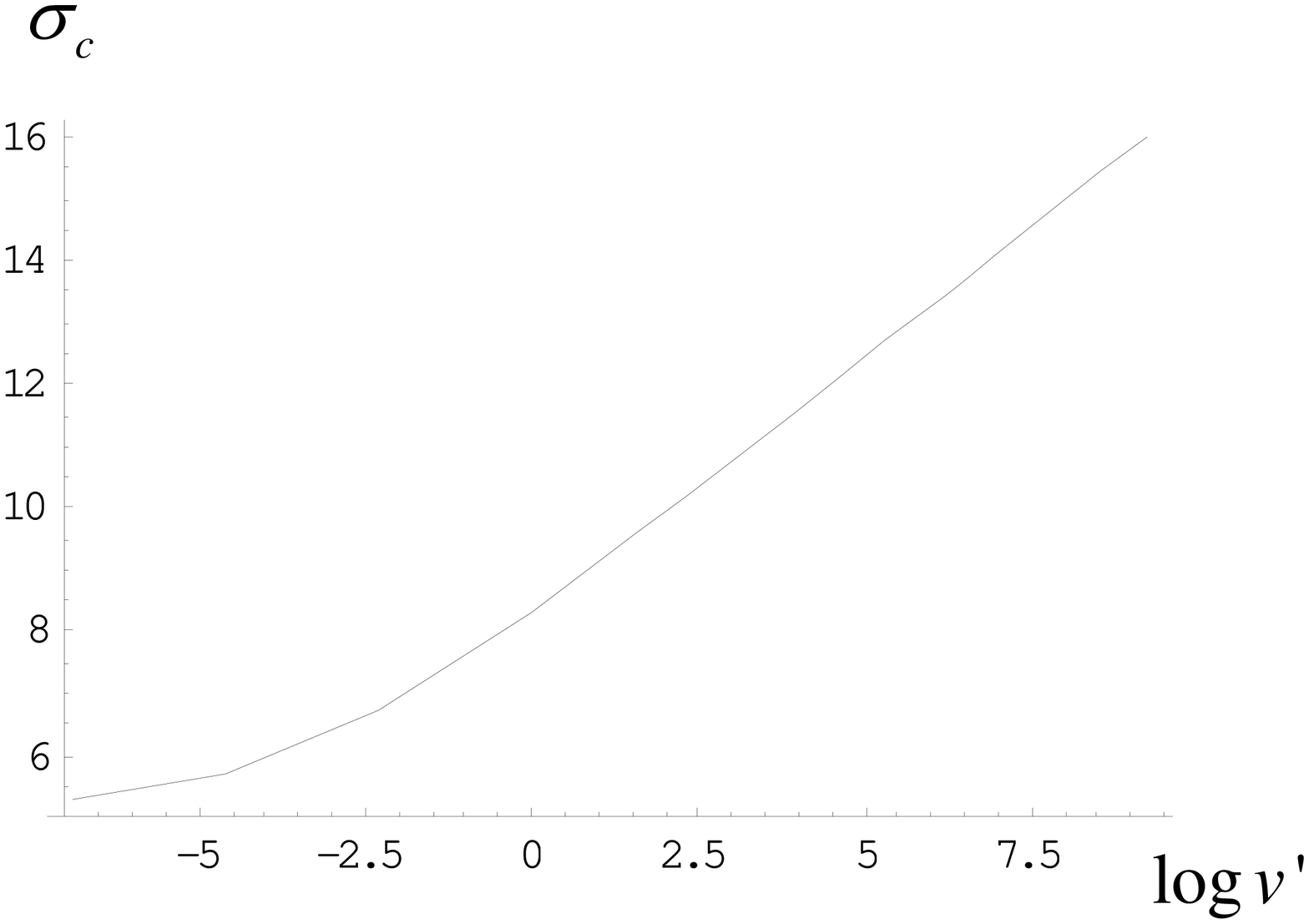} \caption{ Plot of $\sigma_c$ against
$\log v'$ calculated from eq.~(\ref{eq:13}) with $\sigma(t) = e^{v't}$. The parameter
values in the calculation are $\tau=10$, $k_B T=1$, $\ell=1$, and $U=7.7$.  For these
parameter values, $\sigma_s=5.0734.$ }
\label{fig:09} \end{center} \end{minipage}\end{figure}

The behavior with time-dependent $\sigma$ is different from the stationary case.  In our
model, since the flexible joint's curve always has an exponential form, each element is
subjected to $\sigma$ varying exponentially with time; {\it i.e.} $\sigma(t)=e^{v' t}$, in
which $v'$ is a parameter of a positive value.  In such situation, $\sigma$ at the
detaching moment, $\sigma_c$, depends on the rate $v'$ because of the finite relaxation
time in the detachment process; $A$ does not vanish immediately.  We plot $\sigma_c$
against the logarithm of $v'$ in Fig.~\ref{fig:09}, { where we have taken as the
initial condition $A(t=0)=A_0$ where $A_0$ is the stationary value of $A$ for $\sigma=0$
because it is natural to assume that there is no force action on each element before the
peeling process starting (the explict value is $A_0=0.999547$). } Owing to the positive
dependence on $v'$ of $\sigma_c$, the positive slope will appear in the peel force curve
as we usually observe experimentally.

\section{Discussion and Conclusion}

We presented a simple model to describe the peeling system, which is only made of linear
springs and dashpots and solved the system analytically.  With this modeling, the curved
shape of the tape during the peeling is spontaneously determined in terms of the
viscoelastic properties of the adhesive tape.  The advantage of this characteristic is
utilized in the discussion of the two critical peel rates being related with the
viscoelastic properties of the adhesives.

We discussed in $\S$~\ref{sec:positive} the possibility of the peel rate dependence of
the critical stress for detachment; if $\sigma_c$ is constant we can only predict negative
slope of the peel force which is in contradiction with experimental results.  In order to
explain the dependence, we considered the microscopic picture of the element consisting of
small springs repeating attachment and detachment to the substrate.  We admit that there
may be other possibilities for the peel rate dependence of the ``effective" critical
stress, {\it e.g,} the z dependence of the peel line with large amplitude of the
fingering, etc. We have to carefully consider those possibilities theoretically and
experimentally.

\begin{figure}[hbtp]\begin{minipage}[t]{14cm} \begin{center} 
\includegraphics[width=12cm]{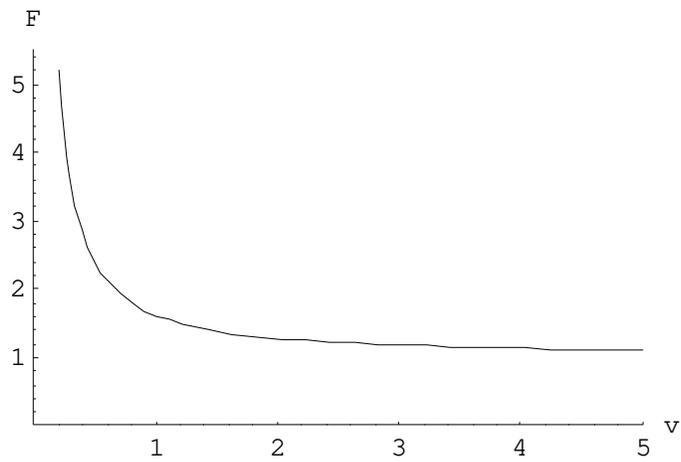}
\caption{The peel force against the peel rate with constant $\sigma_c$(=1) obtained in the
case of linear springs (B) with $K=1$; where the element of the model is a Maxwell element
with $\tau=1$ and $k=1$ } \label{fig:12} \end{center} \end{minipage}\end{figure}

\begin{figure}[hbtp]\begin{minipage}[t]{14cm} \begin{center} 
\includegraphics[width=12cm]{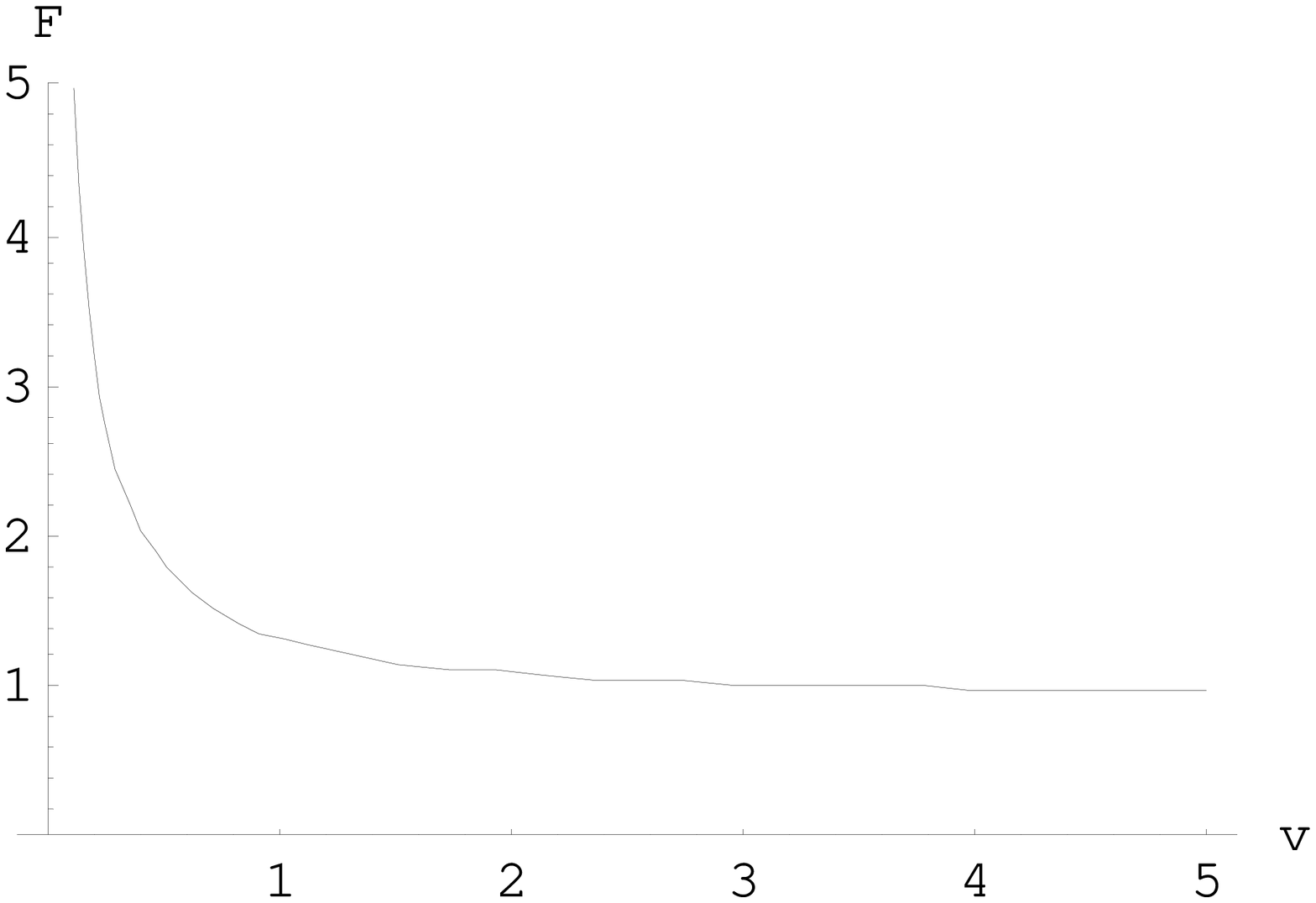} \caption{ The peel force against the peel
rate with constant $\sigma_c$(=1) obtained in the case of bending beam with $K_b=1.25$;
where the element of the model is a Maxwell element with $\tau=1$ and $k=1$, corresponding
to Fig.~\ref{fig:12} for the flexible joints of linear springs (B). }
\label{fig:11} \end{center} \end{minipage}\end{figure}

\section*{Acknowledgment}
The authors thank Mr M. Ohtsuki (Tokyo University), Drs Y. Yamazaki (Waseda University)
and T. Mizuguchi (Osaka Prefecture University) for valuable discussions.  This research
was supported by the 21st Century Center of Excellence Program of the Ministry of
Education, Culture, Sports, Science, and Technology, Japan and by a Grant-in-Aid for
Scientific Research from the Ministry of Education, Culture, Sports, Science and
Technology of Japan.

\appendix

\section{Case of Bending Beam as the Backing Tape} \label{appen:A}

We consider the backing tape as a beam with flexural rigidity $K_b = E_b I_{yy}$, where
$E_b$ represents the modulus of elasticity of the backing tape and $I_{yy}$ is the moment
of the inertia of area and is expressed as $I_{yy} = w \delta^3/12$ for the backing tape
with the width $\omega$ and thickness $\delta$.  The differential equation and its
solution for the case of elastic adhesive has been obtained by
Bikerman~\cite{Bikerman1957} in the following way.  Firstly, the differential equation of
the courved profile of the beam $f(x)$ is given as~\cite{Crandall1959},
\begin{equation} 
K_b \frac{d^4f(x)}{dx^4} = - k f(x), \label{eq:a:22} 
\end{equation} 
where $k$ has the same meaning as in our modeling of the adhesive with springs (A): $k=
E_a \omega / y_0$ for the adhesive with the modulus of elasticity $E_a$, the width
$\omega$ and initial thickness $y_0$.  The boundary conditions are
\begin{eqnarray} 
K_b f''''(0) &=& - \sigma_c, \label{eq:22} \\
f(\infinity) &=& 0 \label{eq:23} 
\end{eqnarray}
and additionally from the moment of force and its balance at $x=0$ 
\begin{equation} f''(0)=0, \label{eq:24} \end{equation} 
\begin{equation} F = K_b f'''(0) . \label{eq:970} \end{equation} 
Solving the differential equation (\ref{eq:a:22}) under the boundary conditions
(\ref{eq:22})-(\ref{eq:24}), we have \begin{equation} f(x)= \frac{\sigma_c}{k} e^{-
x/\alpha_b} \cos (x/\alpha_b) , \end{equation} where $\alpha_b=(4K_b/k)^{1/4}.$ 
The peel force $F$ is then given from eq.~(\ref{eq:970}) as,
\begin{equation}
F=\frac{\alpha}{2} \sigma_c
\end{equation}

In the case of Maxwell elements having coefficients $k$ and $\eta$, we can construct the
equation system and solve it in a similar way. The differential equation for $f(x)$ is
\begin{equation} { \frac{d^4 f(x)}{dx^4} =- 4 \alpha_b^{-4} 
\int_{-\infinity}^{0} e^{t'/\tau} \dot{f}(x-v t') dt' , } \label{eq:969} \end{equation}
\begin{equation} l(x)= \frac{1}{\tau} \int_{-\infinity}^{0} e^{t/\tau}
f(x-v t) dt \end{equation} 
The boundary conditions and the relationship between $F$ and $f$ are the same as (\ref{eq:22})-(\ref{eq:970}).  

Supposing that $f$ has the form $f(x)=A e^{- x / \beta}$ and substituting this form in (\ref{eq:969}), 
we then have an algebraic equation for $\beta$, 
\begin{equation} 
4\beta^{4}  + \frac{\alpha_b^4}{v \tau} \beta +\alpha_b^{4}  =0 
\end{equation}
Note that the real part of the root of this equation must be positive to satisfy the
condition (\ref{eq:23}).
We can find two such solutions; we denote them as $\beta_1^{-1}=\omega + i \gamma$ and
$\beta_2^{-1}=\omega - i \gamma$, where both $\omega$ and $\gamma$ are positive.
Then, we can write $f$ as \begin{equation} f(x)=A e^{-\omega x} \cos(\gamma x + \theta),
\label{eq:25} \end{equation} where $\theta= \cos^{-1} \frac{2 \omega \gamma}{\omega^2
+ \gamma^2} $ ($0 < \theta < \pi$) and 
\begin{equation} 
A= \frac{1}{2\omega\gamma(\omega^2+\gamma^2)} \frac{\sigma_c}{K_b} 
\label{eq:999} 
\end{equation} 
which have been determined from (\ref{eq:22}) and (\ref{eq:24}).
We can calculate the peel force from (\ref{eq:970}), (\ref{eq:25}) and
(\ref{eq:999}), 
\begin{equation} 
F = \frac{1}{2\omega} \sigma_c
\end{equation} 
We plotted the peel force against the peel rate $v$ in Fig.~\ref{fig:11} for $K_b=1.25,
k=1, \tau=1$ and $\sigma_c=1$. { It should be noted that this result is essentially the
same as the result of our modeling with the flexible joints of springs (B) shown in
Fig.~\ref{fig:12} in the following sense; (i) the peel force monotonically decreases with
the peel rate; (ii) the peel force is saturated around some characteristic peel rate,
which is estimated as $(1/\tau) (4K_b/k)^{1/4}$.  (In the case of springs (B) it is
estimated as $(1/\tau) (K/k)^{1/2}$.)  }

\section{Proof of the Form {$f(x)=f(0)\,e^{- x /\beta}$ for 
the Flexible Joints of Linear Spring (B)}} \label{appen:B}

In our model { with the linear springs (B) as the flexible joint } , whatever the linear
combination of linear springs and dashpots, the equation for $f(x)$ necessarily takes the
form as:\begin{equation} { f''(x)=a f(x) + b \left. \dot{f}(x-v t) \right|_{t=0} +
\sum_{i=1}^{M} c_i \int_{-\infinity}^{0} e^{t'/\tau_i} \dot{f}(x-v t') dt'.  }
\label{eq:26} \end{equation} Here, $a$, $b$, and $c_i(i=1,...,M)$ are some non-negative
constants and at least one of them must be nonzero, { $\tau_i$ is the $i$-th relaxation
time of the system, } and the dot over $f$ stands for the differentiation of $f$ with
respect to time.  { For a parallel combination of spring, dashpot and Maxell elements,
there is a direct correspondence between the elements and the terms in the r.h.s. of
(\ref{eq:26}); the first term corresponds to the spring, the second term dashpot, and the
third term Maxell elements.  For more general combination, the expression of (\ref{eq:26})
still holds though there is no such correspondence. }

{ Assume that the solution of (\ref{eq:26}) takes the exponential form of
eq.~(\ref{eq:980}) and substitute this form in (\ref{eq:26}), then } we get an algebraic
equation for $\beta$:\begin{equation} { \beta^{-2} = a + b v \beta^{-1} + \sum_{i=1}^{M}
c_i \frac{v \tau_i}{\beta + v \tau_i} . }
\label{eq:30} \end{equation}

We prove that the algebraic eq. (\ref{eq:30}) has one and only one positive solution.
Let us first introduce a variable $\xi=1/\beta$ and consider a function of $\xi$ defined
on the interval $[0,\infinity]$,\begin{equation} { g(\beta)= \xi^{2} -a - b v \xi -
\sum_{i=1}^{M} c_i \frac{v \tau_i \xi}{1 + v \tau_i \xi}.  }
\label{eq:29} \end{equation}
We can readily find that $g''(\xi) > 0$ for all non-negative $\xi$, i.e., $g$ is a convex
function of $\xi$ on that interval.  On the other hand, for $a=0$, we see that $g(0)=0$
and $g'(0) < 0$ because at least one of $b$ and $c_i$ must be positive. For $a >0 $, we
see that $g(0)<0$.  Noting $g(\infinity)>0$ and remembering that $g$ is a convex function
with respect to $\xi$, we readily see that the equation $g(\xi)=0$ has one and only one
positive solution and accordingly so does the eq. (\ref{eq:30}) for $\beta$; we thus
see the existence of a solution taking the form (\ref{eq:980}).

\end{document}